\documentclass[]{aa}
\usepackage{natbib}
\usepackage{graphicx}
\usepackage{color}
\usepackage{txfonts}

\bibpunct{(}{)}{;}{a}{}{,}

\begin{document}

  \title{Cosmological MHD simulation of a cooling flow cluster}

\author{Yohan Dubois \inst{1} and Romain Teyssier \inst{1}}

\offprints{Y. Dubois}

\institute{Service d'Astrophysique,
CEA/DSM/IRFU/SAp, Centre d'\'Etudes de Saclay, L'Orme des Merisiers,
91191 Gif--sur--Yvette Cedex, France
 \\
\email{ydubois@cea.fr}}
\date{Accepted; Received; in original form;}

\label{firstpage}

\abstract   {Various   observations   of   magnetic  fields   in   the
  intra--cluster medium  (ICM), most of the time  restricted to cluster
  cores, point towards a field strength of  a few $\rm \mu
  G$ (synchrotron radiation from radio relics and radio halos, inverse
  Compton  radiation  in  X-rays   and  Faraday  rotation  measure  of
  polarised  background  sources). Both  the  origin  and the  spatial
  structure of galaxy cluster magnetic  fields are still under debate.
  In particular,  the radial profile  of the magnetic field,  from the
  core  of clusters  to their  outskirts, is   important for
  cosmic ray propagation within the cosmic web.}  {In this letter, we
  highlight  the importance  of  cooling processes  in amplifying  the
  magnetic field  in the core  of galaxy clusters  up to one  order of
  magnitude  above  the  typical  amplification obtained  for  a  purely
  adiabatic evolution.  }  {We  performed a ``zoom'' cosmological
  simulation  of  a 3  keV  cluster,  including  dark matter  and  gas
  dynamics, atomic  cooling, UV heating,  and star formation  using the
  newly developed MHD solver in the AMR code RAMSES.}  {Magnetic field
  amplification     proceeds      mainly     through     gravitational
  contraction. Shearing  motions due to  turbulence provide additional
  amplification  in  the  outskirts  of the  cluster,  while  magnetic
  reconnection during mergers causes magnetic field dissipation in the
  core.  }   {Cooling processes have  a strong impact on  the magnetic
  field structure in the cluster. First,  due to the sharp rise of the
  gas   density  in   the  centre,   gravitational   amplification  is
  significantly  higher, when compared  to the  non--radiative run.
  Second, cooling processes cause shearing motions to be much stronger
  in the core than in  the adiabatic case, leading to additional field
  amplification  and no  significant  magnetic reconnection.   Cooling
  processes  are  therefore  essential for   determining  the
  magnetic field profile in galaxy clusters. }


\authorrunning{Dubois,   Teyssier} 
 
\titlerunning{Cosmological MHD simulation of a cooling flow cluster}

\maketitle

\section{Introduction}

Clusters  of  galaxies are  known  to  be  magnetised (see  review  by
\citealp{govoni&feretti04}).   The existence  of  magnetic fields  has
been  determined either  by  direct methods  like diffuse  synchrotron
radio or  inverse Compton  hard X-ray emission  or by  indirect methods
like Faraday  rotation measures (RM). They  all suggest that  $ \mu G$
fields lie in central regions of  galaxy clusters up to several $10 \,
\rm \mu G$ magnetic fields  in large cooling flows clusters.  Magnetic
field strength  can differ  from one method  to the next  Direct methods
usually capture large--scale  fields averaged over large volumes, while
RM are derived from the  analysis of background point sources and are
thus sensitive to small--scale variations (cold filaments, shear flows,
shocks, galaxy stripping, galaxy winds, etc.). To shed light on
the  magnetic topology  found in  cosmic  structures, it  is of  great
interest to  perform direct, self-consistent  numerical simulations of
galaxy  clusters.   Magnetic fields  in  clusters  are  also 
important for determining  the  deflection angle  of ultra-high  energy
cosmic rays, since they probably host the source of these cosmic rays
(see \citealp{lemoine05, kotera&lemoine07, globusetal07}).

Simulations  of  galaxy  clusters   with  magnetic  fields  have  been
performed   using   smooth    particle   hydrodynamics   (SPH)   codes
\citep{dolagetal99, dolag00, dolagetal05},        grid--based       codes
\citep{roettigeretal99,  miniatietal01,  sigletal04,  asaietal07},  and
adaptive mesh refinement (AMR) codes \citep{bruggenetal05}, using both
cosmological     simulations     \citep{dolagetal99,    miniatietal01,
sigletal04,      dolagetal05} and      idealised      simulations
\citep{roettigeretal99,  asaietal07}.  In this  letter, we  report the
first cosmological  simulation with AMR to  include atomic cooling,
UV heating,  and star formation physics,  with a full  treatment of the
ideal MHD equations.  We have also performed a reference adiabatic run
to  compare our results with previous works  and to point out
the differences with the radiative case.

\section{Simulations}

We  performed  a  ``zoom'' cosmological  simulation  of a  galaxy
cluster  using  the   AMR  code  RAMSES  (\citealp{teyssier02}).   Gas
dynamics was computed using  a second--order unsplit Godunov scheme for
the ideal  MHD equations \citep{teyssieretal06,  fromangetal06}, while
collisionless dark matter particles are evolved using a particle--mesh
solver. Gas cooling and heating were taken into account as source terms
in  the  energy  equation.   The  cooling and  heating  functions  were
computed   for   a   primordial   H   and   He   plasma,   using   the
\cite{haardt&madau96} background model.  Radiative losses lead to the
formation of high density,  (low temperature) regions, where stars are
allowed to  form according to a  Schmidt law: $\dot  \rho_* = \epsilon
\rho /  t_{\rm ff}$ if  $\rho>\rho_0$. The density threshold  for star
formation  was  set to  $\rho_0=10^5  \Omega_b  \rho_c(z)$.  The  star
formation  efficiency  was  set  to  $\epsilon=5$\%.   The  simulation
comoving box length was chosen equal  to $80 \, \rm h^{-1}$ Mpc with a
$\Lambda$CDM  cosmology   with  $\Omega_m=0.3$,  $\Omega_\Lambda=0.7$,
$\Omega_b=0.045$,  $H_0=70 \,  \rm km.s^{-1}.Mpc^{-1}$,  and 
  $\sigma_8=0.9$. A  spherical region of radius $12.5  \, \rm h^{-1}$
Mpc around  our simulated cluster  was defined as  our high-resolution
region, with an  effective resolution of $512^3$. A  coarser grid with
an effective resolution of $256^3$ was  used to cover the inner $40 \,
\rm h^{-1}$ Mpc, and finally an  even coarser $128^3$ grid was used to
cover the whole box.  The mass of dark matter particles on each coarse
grid are $2.9\times 10^{10} \, \rm M_{\odot}$, $3.6\times
10^{9}  \, \rm  M_{\odot}$, and  $4.5\times 10^{8}  \,  \rm M_{\odot}$.
Only the finest grid was allowed to trigger new refinements during the
course  of the  simulation, up  to 7  additional levels  of  AMR cells
leading  to  a maximum  resolution  of  $1.2 \,  \rm
    h^{-1} kpc$.  We used a quasi--Lagrangian criterion: each cell is
individually refined if the number of dark matter particles exceeds 8,
or if  the baryonic mass  exceeds 8 times the  initial high-resolution
mass resolution.  We solved the  full set of ideal MHD equations using
a new scheme based on a Godunov implementation of constrained transport
and presented in \cite{teyssieretal06} and \cite{fromangetal06} and we
used  the  HLLD  Riemann   solver  from  \cite{miyoshi05}.   The  {\it
  comoving} magnetic field was set initially to a constant value, $B_z
\simeq 10^{-11}$  G, as  suggested in \cite{dolagetal05},  to reproduce
the $\rm  \mu G$ fields in  cluster cores.  With  these parameters in
the course of the simulation, the plasma $\beta=P_{gas}/P_{mag}$ never
decreased below 1000:  the dynamical effect of the  magnetic field can
therefore  be  considered as  negligible,  even  in  the core  of  our
simulated cluster.

\section{Results}

\begin{figure}
\centering{\resizebox*{!}{5.5cm}{\includegraphics{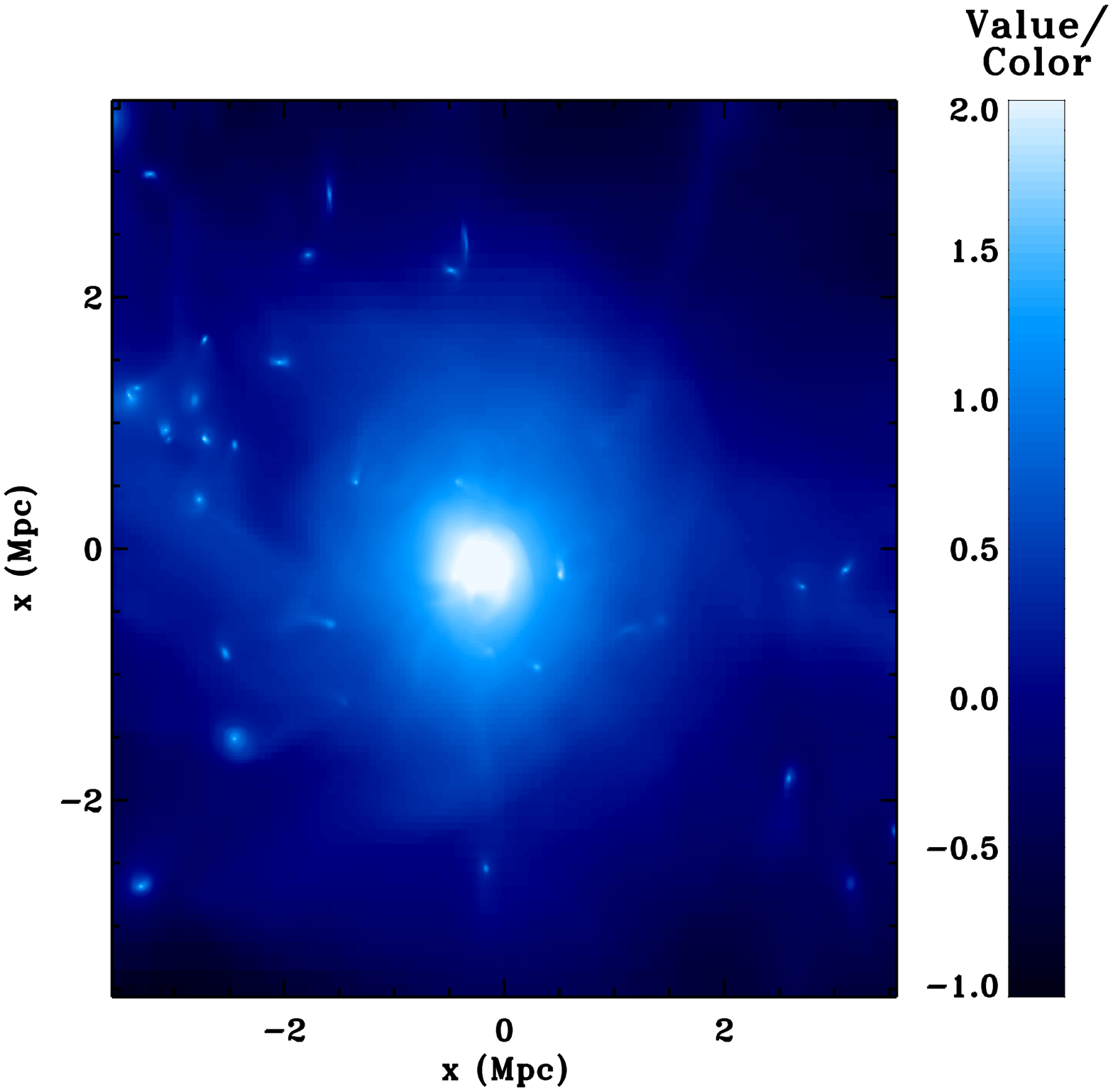}}}
\centering{\resizebox*{!}{5.5cm}{\includegraphics{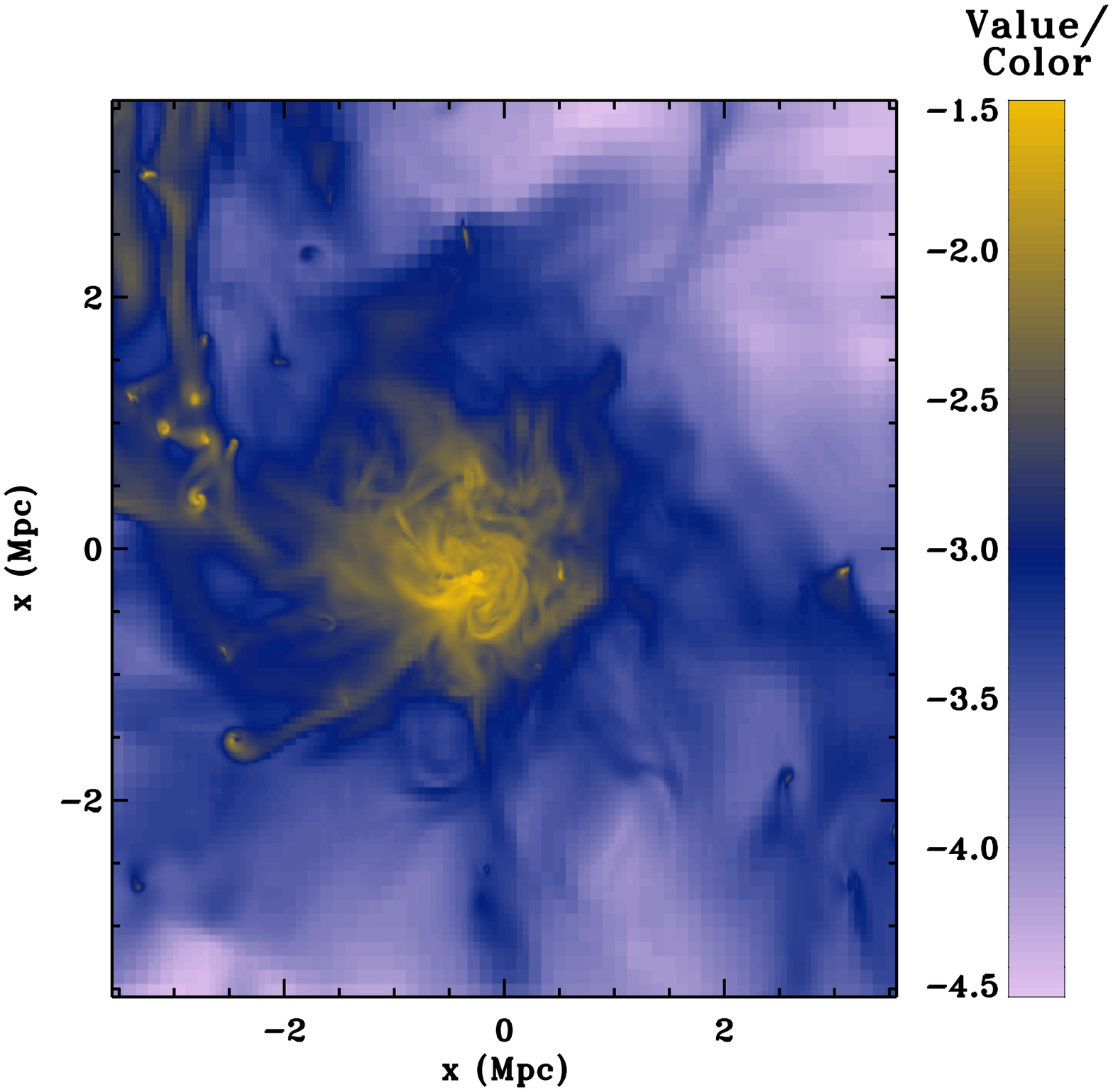}}}
\caption{Logarithm of the  column density map of the  gas in units of the mean baryons density (upper panel), and of  the mass--averaged magnetic  field amplitude in units of $\rm \mu G$ (bottom panel) at $z=0$. }
\label{rho_snap}
\end{figure}

{Figure~\ref{rho_snap} shows the column density distribution of the gas
and the mean magnetic field for the cooling run at  $z=0$.  Magnetic  field  amplitudes are strongly correlated with  the
density distribution  with mass-averaged values of $B  \sim 10^{-1} \,
\rm \mu G$  in the cluster core, galaxies with $B  \sim 10^{-2} \, \rm
\mu  G$ in  satellite clumps,  and $B  \sim 10^{-3}  \, \rm  \mu  G$ in
filaments.  We define the virial mass as $M_{200}=200 \times 4 \pi / 3
\rho_{c}  R_{200}^3 \, ,$  where $\rho_{c}$  is the  critical density.
For the  adiabatic simulation, we found the
following  properties for our cluster  at $z=0$:  $R_{200}^{ad} \simeq  1  \,  \rm h^{-1}  Mpc$,
$M_{200}^{ad} \simeq  2.7 \times  10^{14}\, \rm h^{-1}  M_{\odot}$, and
$T_{X}^{ad} \simeq 3.4\, \rm keV$.  In the radiative case, we obtained
$R_{200} \simeq  1.1 \,  \rm h^{-1} Mpc$,  $M_{200} \simeq  3.5 \times
10^{14}\, \rm h^{-1} M_{\odot}$, and $T_{X} \simeq 5.1\, \rm keV$.  The
magnetic  field amplification  of a  collapsing three--dimensional gas
sphere with infinite conductivity  is given by $B \propto \rho^{2/3}$,
for the magnetic flux to be conserved.  Thus a subsequent increase (or
decrease)  in  the  magnetic  field  with  respect   to  this  purely
gravitational  amplification  should   reveal  other  amplification  or
dissipation mechanisms.   Figure~\ref{bvsrho} shows the mass--weighted
histogram  of the  radiative simulation  in  the $\rho$-$|B|$
plane, within 2 virial radii around the cluster.
 For densities lower than $10^4 \bar \rho$, the
magnetic field amplification is one order of magnitude higher than for
pure gravitational  compression.  As discussed  in \cite{dolagetal05},
this is likely due to  shearing motions in the cluster atmosphere, caused by  turbulence  and  frequent   mergers.   At  higher  densities,  the
radiative  run  diverges  strongly   from  the  adiabatic  case.   The
gravitational compression due to  the cooling flow provides additional
field amplification in  the high--density tail. As seen below,
cooling also provides a sustained turbulent regime in the core and the
corresponding additional field  amplification. Based on the Zel'dovich
approximation of  gravitational dynamics, \cite{king&cole06} predict
that  a cosmological  magnetic field  should  evolve as  a $B  \propto
\rho^{0.87}$, due to anisotropic  collapse in a Gaussian random field.
This compares favourably  with the low--density part  of our simulation
(see Fig.~\ref{bvsrho}).

\begin{figure}
\centering{\resizebox*{!}{5.5cm}{\includegraphics{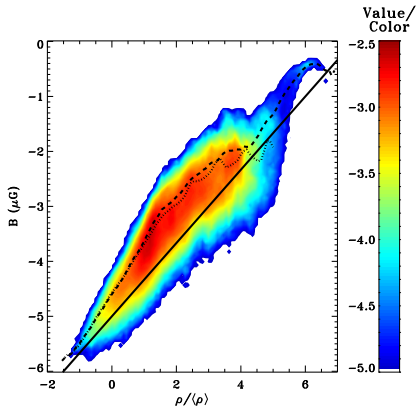}}}
\caption{Histogram  of the  mass fraction  for  the cooling  run as  a
  function  of the normalised  density and  the magnetic  amplitude at
  $z=0$.   The  black   solid  line   is  the   $\rho^{2/3}$  collapse
  amplification,  the averaged  magnetic field  as a  function  of the
  normalised  density  for the  cooling  case  (dashed  line) and  the
  adiabatic case (dotted line) are also plotted.}
\label{bvsrho}
\end{figure}

At higher density,  the situation
is more complex in the core of our simulated cluster.  In the  adiabatic case, the magnetic field amplitude
decreases  {\it below} the  expected value  for pure  compression.  The
mean magnetic  field is of  the order of  $10^{-2} \, \rm \mu  G$, far
below  the typical  values of  magnetic  field amplitude  in
observed cluster  cores.  It  is also apparent  in the  magnetic field
profile plotted  in Fig.~\ref{bvsr},  for which a  dip in  the field
strength is visible  in the cluster core. While shear  flows are able to
sustain  additional  magnetic  amplification  in  the outer  parts  of  the
cluster  $  r  >  150  \,  \rm h^{-1}  kpc$,  we  find the opposite behaviour in the cluster core where magnetic field is  {\it
  dissipated}.    According to   
\cite{roettigeretal99}, this  effect   stems from magnetic reconnection
occurring  during merger  events.   Since we  are  not considering  any
microscopic process here and since the MHD turbulence scale is far
  below  our numerical resolution,  this dissipation  is  caused by our
numerical  scheme that  captures the  weak solution  of the  ideal MHD
equations. Although  magnetic reconnection probably  occurs in nature
within  converging flows, the  exact amplitude  of the  dissipation is
likely  to  depend on  the  microphysics.   In  the present  numerical
approach, results should depend strongly on the spatial resolution and
on the numerical scheme used.  It is however interesting to analyse the effect  of  cooling in
this  respect.   As  can be  seen  in
  Fig.~\ref{bvsr},  magnetic  dissipation in  the  cluster core  is
  suppressed
  and magnetic amplification now proceeds  in the same way as in the
  outer  parts, with gravitational  compression and  shearing motions.
  Only in the  very centre (below $3 \, \rm h^{-1}  kpc$, close to the
  resolution limit) do we see magnetic dissipation again.

\begin{figure}
\centering{\resizebox*{!}{5.5cm}{\includegraphics{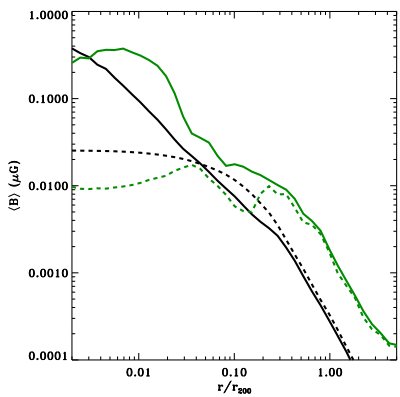}}}
\caption{Mean  magnetic   field  as  a  function   of  the  normalised
radius. The  black lines are the  $\rho^{2/3}$ collapse amplification,
the green  lines denote the run values for  the adiabatic
(dotted) and cooling (solid) runs at $z=0$.}
\label{bvsr}
\end{figure}

\begin{figure}
\centering{\resizebox*{!}{5.5cm}{\includegraphics{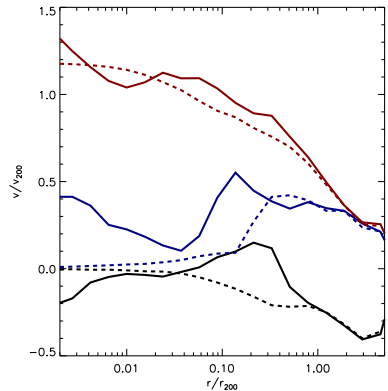}}}
\caption{Mean  radial  velocity  (black), radial  velocity  dispersion
(blue), and sound speed (red) for the adiabatic (solid) and the cooling
(dashed) runs at $z=0$ in units of $V_{200}$ for the adiabatic case.}
\label{turb}
\end{figure}

To  illustrate  this  point  further,  we  show  in  Fig.~\ref{turb}
velocity  profiles in  the adiabatic  and  in the  cooling cases.   The
radial velocity dispersion is a  signature of turbulent motions, so it
is no surprise to see  that a strong velocity dispersion at a given
radius corresponds  to an  excess of field  amplification at  the same
radius (see  Fig.~\ref{bvsr}). In the adiabatic  case, turbulence is
dissipated  in the  core,  and magnetic  dissipation  occurs. In  the
cooling case,  gravitational contraction resumes, as  well as shearing
motions,  so that  magnetic amplification  is now  more active  in the
core.  The RM  maps at $z=0$ of  the adiabatic simulation  and the cooling
simulation  shown  in  Fig.~\ref{RMmap}  differ strongly.   In  the
adiabatic case, we  obtain RM values of $30 \,  \rm rad.m^{-2}$ in the
cluster centre, whereas  in the radiative case, we  reach $1000 \, \rm
rad.m^{-2}$  in the  very centre  ($r<10 \,  \rm h^{-1}  kpc$)  and 
obtain RM values  of $100 \, \rm rad.m^{-2}$ in  the core. The RM results
of the cooling simulation  are marginally consistent with the \cite{clarkeetal01}
sample ($200\, \rm rad.m^{-2}$) and also with the maximum values found
in  the cluster  sample of  \cite{tayloretal02}  (up to  $1800 \,  \rm
rad.m^{-2}$ for  the hot gas cluster).  The magnetic
  field in the  outer part of our  cluster is on the weak  side of the
  \cite{clarkeetal01}  sample.  It  could result  from  differences in
  cluster  mass  or from  our  somewhat  arbitrary  choice of  initial
  magnetic  field. Note  that we  can increase  the
  initial amplitude  up to  $10$  times while remaining in the linear regime ($\beta \ge
  10$), so that all our results can be re-normalised by this factor.

\begin{figure}
\centering{\resizebox*{!}{5.5cm}{\includegraphics{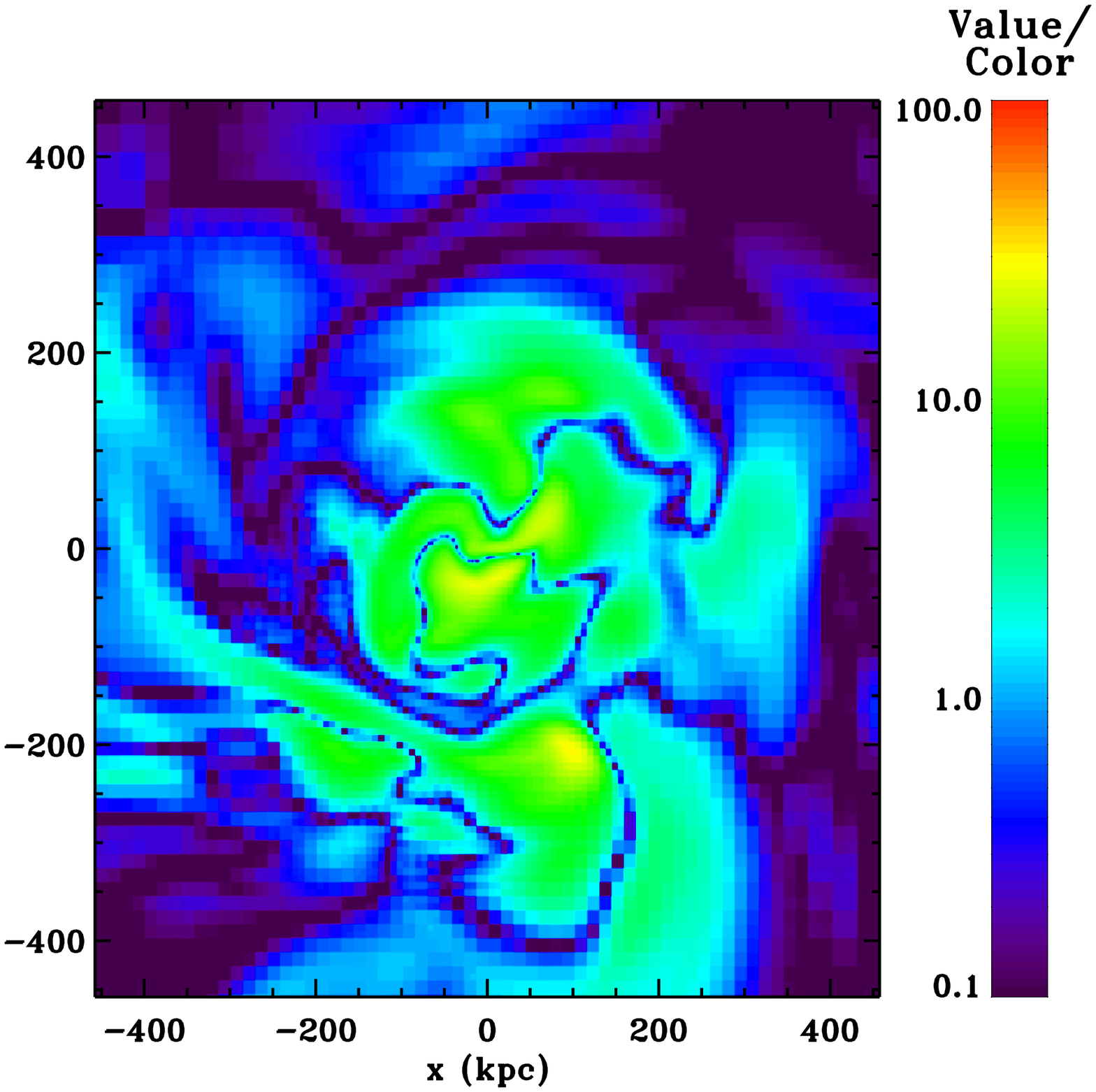}}}
\centering{\resizebox*{!}{5.5cm}{\includegraphics{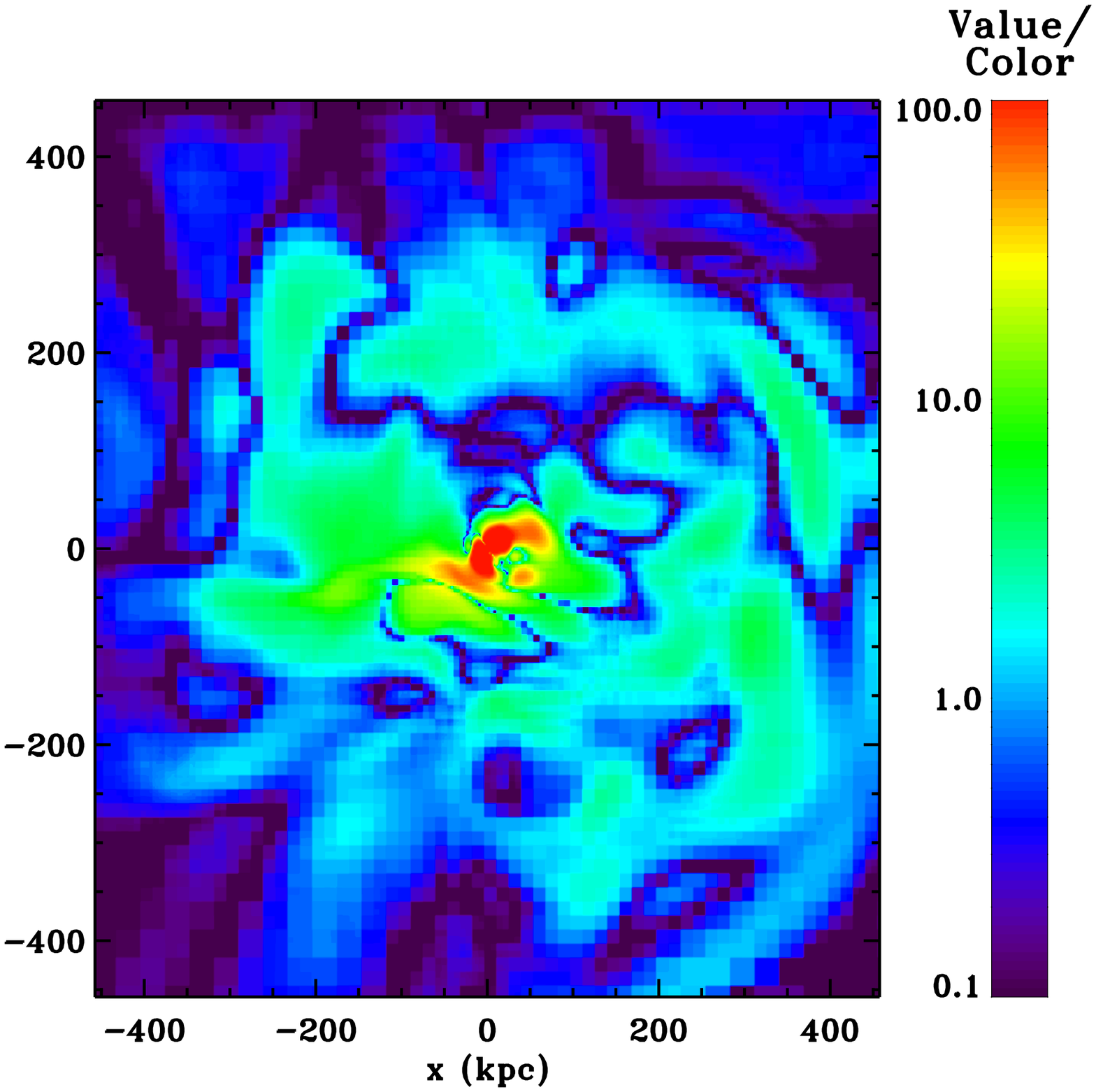}}}
\caption{$\vert$RM$\vert$ maps  colour scale of  the cluster core for  the adiabatic
run (left panel) and the cooling run (right panel) at $z=0$.}
\label{RMmap}
\end{figure}

\section{Conclusion and discussion}

There are noticeable differences in the magnetic field characteristics
between  a galaxy  cluster with  an adiabatic  evolution and  a galaxy
cluster  with radiative  cooling: the  average magnetic  field  in the
cluster core is  significantly higher when a cooling  flow is present,
due  to  additional  gravitational  compression  but also   an
increased level  of turbulence in  the core--driving  shearing motions.
The main  consequence is that Faraday RM simulated maps
agree more with  observations in the cooling case, if the
initial comoving magnetic field  value is taken to be $10^{-11}$
G.  In low--density regions, however, the magnetic  field evolution in
the radiative  run is very  close to the adiabatic  case since
  dynamical properties of the gas turbulence are nearly the same. Note
  that the turbulent  cascade is far from being  fully resolved in our
  simulation, especially outside the cluster core where our resolution
  barely reaches  $10 \, \rm h^{-1}  kpc $, a much larger scale than
  the     dissipation     scale     of    MHD     turbulence     (e.g.
  \citealp{brunetti&lazarian07,  jones07}).   For  these  reasons  our
  results  should  only  be  considered  as lower  estimates  for  the
  magnetic field strength.

 As was  already discussed  in
\cite{roettigeretal99}  in  the context  of  adiabatic simulations  of
idealised  mergers, magnetic reconnection might be responsible for field
dissipation  in  the cluster  core. Since  we  are dealing  with  ideal MHD,  magnetic
dissipation  occurs at  the numerical  level,  so that  we should  be
affected  to  some  extent  by  the effect  of  numerical  resolution.
Moreover,  this  underlines  the  importance  of  the  choice  of  the
numerical  code used,  especially when  one considers  the fundamental
differences  between  grid--based and  particle--based codes.
Using  an MHD  version  of GADGET,  \cite{dolagetal05}  find a much
 larger magnetic  field strength  in  their
adiabatic  run for a  subset of
their simulated particles. They report  mean field values one order of
magnitude higher than median field values, apparently contradicting
 our present result.  Magnetic dissipation  therefore appears  much
less efficient in the SPH  case. The Lagrangian nature of this
  technique is radically different  from grid-based codes.  Since both
  codes  lead   to  different  magnetic  diffusion   terms,  a  proper
  comparison between numerical techniques  on this stake would be very
  helpful  to  the  community.    Using  the  grid--based  code  ZEUS,
\cite{roettigeretal99}  report  very  similar results  to  ours,  with
strong field dissipation occurring in the  converging part of the flow. One
interesting  outcome of  the present  work is  that  radiative cooling
drastically  changes  the   effect  of  magnetic  reconnection,  since
turbulence  and gravitational  compression  easily counterbalance  the
associated dissipation.

Nevertheless, this  large additional amount of magnetic  energy in the
core  of  cooling  clusters  is  a crucial  step  in  determining  the
structure  of  the  cosmological  magnetic  field.  It  has  a  direct
consequence  on the  propagation of  high--energy cosmic  rays in  the
universe.   Since we have  no direct  observations of  magnetic fields
outside of cluster cores,  only cosmological numerical simulations can
address  this problem.  Their  weakness is  that the  initial magnetic
field  value must  be  normalised a  posteriori  in order  to fit  the
observed  values.  We have  shown that,  for the  same seed  field, the
final magnetic field  strength in a cooling cluster  core is one order
of magnitude more than in the adiabatic case.  Cooling processes are
therefore   important if  one wants  to describe  the proper
evolution of magnetic fields in the Universe.

\bibliographystyle{aa}
\bibliography{author}

\end{document}